\documentclass[10pt,twocolumn,english,prd,superscriptaddress,nofootinbib,preprintnumbers,showpacs,floatfix]{revtex4-2}

\usepackage[utf8]{inputenc}
\usepackage{amsmath}
\usepackage{amsfonts}
\usepackage{amssymb}
\usepackage{float}
\usepackage{graphicx} 
\usepackage[caption=false]{subfig}
\graphicspath{ {figures/} }
\usepackage[usenames,dvipsnames]{xcolor}
\usepackage{hyperref}   
\hypersetup{
	colorlinks=true, 
	citecolor=MidnightBlue,
	linkcolor=MidnightBlue,
	urlcolor=Cyan}
\usepackage{tikz}
\usepackage{subcaption}
\usepackage{orcidlink}
\usepackage{verbatim} 
\usepackage{soul} 
\definecolor{purple}{rgb}{1,0,1}
\definecolor{lime}{HTML}{A6CE39} 

\newcommand{\orcidicon}{%
	\begin{tikzpicture}
		\draw[lime, fill=lime] (0,0) 
		circle [radius=0.16] 
		node[white] {{\fontfamily{qag}\selectfont \tiny ID}};
		\draw[white, fill=white] (-0.0625,0.095) 
		circle [radius=0.007];
	\end{tikzpicture}	\hspace{-2mm}
}
\newcommand\orcidEdnaldo{{\href{https://orcid.org/0000-0001-7230-3666}{\orcidicon}}}
\newcommand\orcidFrancisco{{\href{https://orcid.org/0000-0002-9388-8373}{\orcidicon}}}
\newcommand\orcidManuel{{\href{https://orcid.org/0000-0001-8586-0285}{\orcidicon}}}
\newcommand\orcidTarciso{{\href{https://orcid.org/0009-0007-0450-2672}{\orcidicon}}}
\newcommand\orcidHenrique{{\href{https://orcid.org/0000-0001-7565-4277}{\orcidicon}}}
\newcommand\orcidLuis{{\href{https://orcid.org/0009-0009-4322-6484}{\orcidicon}}}

\newcommand\orcidJorde{{\href{https://orcid.org/0009-0001-3344-2986}{\orcidicon}}}

\newcommand\orcidDiego{{\href{https://orcid.org/0000-0003-3984-9864}{\orcidicon}}}
\begin{document}
	
\title{Scalar--Electromagnetic Couplings as Source of Deformed Black Hole:\\ From Shadows to Thermodynamic Topology}

\author{Ednaldo L. B. Junior\orcidEdnaldo\!\!} \email{ednaldobarrosjr@gmail.com}
\affiliation{Faculdade de F\'{i}sica, Universidade Federal do Pará, Campus Universitário de Tucuruí, CEP: 68464-000, Tucuruí, Pará, Brazil}
\affiliation{Programa de P\'{o}s-Gradua\c{c}\~{a}o em F\'{i}sica, Universidade Federal do Sul e Sudeste do Par\'{a}, 68500-000, Marab\'{a}, Par\'{a}, Brazill}
	
\author{Jos\'{e} Tarciso S. S. Junior\orcidTarciso\!\!}
\email{tarcisojunior17@gmail.com}
\affiliation{Faculdade de F\'{\i}sica, Programa de P\'{o}s-Gradua\c{c}\~{a}o em F\'isica, Universidade Federal do Par\'{a},  66075-110, Bel\'{e}m, Par\'{a}, Brazil}
	
\author{Francisco S. N. Lobo\orcidFrancisco\!\!} \email{fslobo@ciencias.ulisboa.pt}
\affiliation{Instituto de Astrof\'{i}sica e Ci\^{e}ncias do Espa\c{c}o, Faculdade de Ci\^{e}ncias da Universidade de Lisboa, Edifício C8, Campo Grande, P-1749-016 Lisbon, Portugal}
\affiliation{Departamento de F\'{i}sica, Faculdade de Ci\^{e}ncias da Universidade de Lisboa, Edif\'{i}cio C8, Campo Grande, P-1749-016 Lisbon, Portugal}
	
\author{Jorde A. A. Ramos\orcidJorde\!\!}
\email{jordealves@ufpa.br}
\affiliation{Faculdade de F\'{\i}sica, Programa de P\'{o}s-Gradua\c{c}\~{a}o em 
F\'isica, Universidade Federal do Par\'{a},  66075-110, Bel\'{e}m, Par\'{a}, Brazil}
	
\author{Manuel E. Rodrigues\orcidManuel\!\!}
\email{esialg@gmail.com}
\affiliation{Faculdade de F\'{\i}sica, Programa de P\'{o}s-Gradua\c{c}\~{a}o em F\'isica, Universidade Federal do 
Par\'{a},  66075-110, Bel\'{e}m, Par\'{a}, Brazil}
\affiliation{Faculdade de Ci\^{e}ncias Exatas e Tecnologia, 
Universidade Federal do Par\'{a}\\
Campus Universit\'{a}rio de Abaetetuba, 68440-000, Abaetetuba, Par\'{a}, Brazil}
	
\author{Diego Rubiera-Garcia\orcidDiego\!\!} 
\email{ drubiera@ucm.es}
\affiliation{Departamento de Física Téorica and IPARCOS, Universidad Complutense de Madrid, E-28040 Madrid, Spain}

\author{Luís F. Dias da Silva\orcidLuis\!\!} 
\email{fc53497@alunos.fc.ul.pt}
\affiliation{Instituto de Astrof\'{i}sica e Ci\^{e}ncias do Espa\c{c}o, Faculdade de Ci\^{e}ncias da Universidade de Lisboa, Edifício C8, Campo Grande, P-1749-016 Lisbon, Portugal}
	
\author{Henrique A. Vieira\orcidHenrique\!\!} \email{henriquefisica2017@gmail.com}
\affiliation{Faculdade de F\'{i}sica, Programa de P\'{o}s-Gradua\c{c}\~{a}o em F\'{i}sica, Universidade Federal do Par\'{a}, 66075-110, Bel\'{e}m, Par\'{a}, Brazil}

\begin{abstract}
We reconstruct a static and spherically symmetric black hole geometry originally proposed as an effective metric by identifying a consistent matter source derived from a fundamental action. The space-time is supported by a magnetically charged nonlinear electrodynamics (NED) field non-minimally coupled to a scalar field. Dimensional consistency reduces the parameter space to a single magnetic charge, and the inverse construction formalism yields a one-parameter family of electromagnetic Lagrangians $\mathcal{L}(F)=F^{n+1}/(n+1)$, encompassing both linear and nonlinear electrodynamics. We analyze the horizon structure and determine the critical magnetic charge separating black hole and horizonless configurations. The photon sphere and the corresponding shadow radius are computed, and observational bounds from the Event Horizon Telescope for Sagittarius~A* constrain the allowed range of the magnetic charge. In the extended phase space thermodynamics, the solution satisfies the first law and the Smarr relation, exhibits a Hawking--Page phase transition, and presents a single change in stability without van der Waals--type critical behavior. We also investigate the topological properties of both the photon sphere and the thermodynamic parameter space. The photon sphere carries a total topological charge $Q_{\text{tot}}=-1$, while the thermodynamic vector field yields a global winding number $W=0$, placing the solution in the same topological class as the one of the Reissner--Nordstr\"om black hole. We finally discuss the versatility of this non-minimal coupling framework in both providing theoretical support to previously introduced solution and also to connect them to observational settings within strong-field gravity. 	
\end{abstract}

\date{\today}

\maketitle

\section{Introduction}

Since its formulation, the General Theory of Relativity (GR) \cite{Einstein1905} has established itself as the successful description of the gravitational interaction, predicting a wide range of phenomena with remarkable accuracy, from the dynamics of the solar system to cosmological evolution \cite{Will:2018mcj,Will:2014zpa}. One of its most intriguing predictions is the existence of black holes, defined by the presence of an event horizon. Long considered as purely theoretical constructs, the detection of gravitational waves and the imaging of black holes from accretion disks, have provided us with convincing arguments in favor of their existence. Regarding black hole imaging (``shadows"), in a series of groundbreaking results, the Event Horizon Telescope (EHT) Collaboration, a global very long baseline interferometry array, unveiled the first direct images of the supermassive black holes at the center of the galaxy M87 \cite{EventHorizonTelescope:2019ggy} and, subsequently, of our own galaxy, the Milky Way, whose core is dubbed as Sagittarius A* (Sgr A$^*$) \cite{EventHorizonTelescope:2022wkp}. 

The imaging of M87$^*$ and Sgr A$^*$ provides a laboratory to test modified theories of gravity and constrain deviations from the Schwarzschild and Kerr metrics \cite{Vagnozzi:2022moj}. In fact, at the theoretical level GR suffers from several inconsistencies, most notably the prediction of space-time singularities \cite{Hawking:1973uf}, regions in which trajectories of particles come to an abrupt end. Singularities are frequently interpreted as signaling the limit of validity of the classical theory. This way, the search for a description of ultra-compact objects that is free from these pathologies is one of the main challenges of contemporary theoretical physics \cite{Ansoldi:2008jw, Carballo-Rubio:2019fnb}. A promising approach to bypass or even eliminate central singularities is the coupling of gravity to more complex matter fields, such as nonlinear electrodynamics (NED) \cite{Bronnikov:2000vy, Ayon-Beato:1998hmi,Fan:2016hvf} and scalar fields \cite{Faraoni:2004pi, Herdeiro:2015waa}, which emerge naturally in low-energy scenarios of fundamental theories, including string theory \cite{Polchinski:1998, Green:1987}.
	
On the other hand, non-minimal couplings between scalar fields and the electromagnetic sector arise naturally in scalar-tensor theories and play a central role in mechanisms of spontaneous scalarization \cite{Damour:1993hw, Doneva:2022ewd}. In this framework, the scalar field becomes non-trivial in the presence of strong electromagnetic fields, modifying the space-time geometry without introducing new independent charges. This approach can be used to construct black hole solutions that are indistinguishable from their GR counterparts in weak-field regimes but exhibit significant deviations in strong-field scenarios, potentially leaving observable imprints in electromagnetic and gravitational wave signatures \cite{Herdeiro:2015waa, Berti:2015itd}. 

Recently, two studies demonstrated the versatility of this mechanism in complementary directions. In Ref.~\cite{Cordeiro:2025ydg} it was shown that the well known Bardeen regular black hole can be reinterpreted as a solution of GR coupled to a scalar field and a linear electromagnetic field. In a subsequent work, the same framework was extended to black bounce space-times \cite{Cordeiro:2025ivw}, encompassing traversable wormholes and regular black holes as limiting cases. In these studies, a power-law form for the derivative of the electromagnetic Lagrangian, \(\mathcal{L}_F(r) \sim F^n\) is imposed, then the coupling function \(W(\varphi)\) is reconstructed, and furthermore it is demonstrated that these nontrivial geometries can be supported by a linear electromagnetic field. These findings highlight the power of non-minimal couplings to reveal alternative and often simpler matter descriptions for existing effective geometries.
	
Another significant area in GR is the thermodynamics of black holes. Following the works of Bekenstein, Hawking, and Bardeen \cite{Bekenstein:1973ur, Hawking:1975vcx, Bardeen:1973gs}, black holes are now understood as thermodynamic systems possessing temperature, entropy and so on, and satisfying laws analogous to those of ordinary thermodynamics. This thermodynamic interpretation provides a powerful tool to probe the stability of black hole solutions through the analysis of heat capacity and free energy \cite{Davies:1977bgr}. More recently, the extended phase space formalism, in which the cosmological constant is interpreted as a thermodynamic pressure and its conjugate as a thermodynamic volume, has revealed a phase structure for black holes in anti-de Sitter (AdS) spacetimes \cite{Kastor:2009wy, Dolan:2011xt, Kubiznak:2012wp}. Within this framework, black holes exhibit critical behavior and phase transitions analogous to those of van der Waals fluids.

A notable feature in asymptotically Anti-de Sitter (AdS) backgrounds is the Hawking--Page phase transition \cite{Hawking:1982dh}, a first-order transition between thermal AdS space, dominated by radiation, and a stable large black hole phase. At low temperatures, thermal AdS is thermodynamically preferred, while above a critical temperature the large black hole dominates, indicating a discontinuous change in the free energy.
Within the AdS/CFT correspondence, this transition is interpreted as the dual of the confinement--deconfinement phase transition in large \(N\) gauge theories \cite{Witten:1998zw, Witten:1998qj}. This provides a direct link between black hole thermodynamics and the phase structure of strongly coupled quantum field theories.

Complementary to thermodynamic studies, a topological perspective on black hole solutions has gained increasing attention. This line of inquiry was pioneered by Cunha and collaborators. In a series of works \cite{Cunha:2017qtt, Cunha:2020azh, Xavier:2024iwr}, they demonstrated that axisymmetric, stationary ultracompact solutions satisfying the null energy condition must possess an even number of light rings, with at least one of them being stable. Their analysis, grounded in topological arguments based on the Brouwer degree of a continuous map, revealed that light rings emerge as topological defects in an appropriately constructed vector field, independent of the specific gravitational theory under consideration. Building upon this foundation, subsequent developments have shown that both the photon sphere and the thermodynamic parameter space of black holes can be characterized through topological invariants in a unified manner \cite{Wei:2020rbh, Wei:2022dzw}. In this approach, the photon sphere, corresponding to the last unstable circular orbit for massless particles, appears as a topological defect carrying a charge that distinguishes black hole space-times from horizonless compact objects. Similarly, the zeros of a vector field constructed from the generalized off-shell free energy classify black hole solutions according to their thermodynamic stability and phase structure.

Within these contexts, recently a new static and spherically symmetric black hole solution was proposed by Khosravipoor et al.~\cite{Khosravipoor:2023jsl}. This geometry introduces deformation parameters that modify the metric function near the origin while maintaining asymptotic (A)dS behavior. Although originally constructed as an effective space-time supported by a prescribed energy density, its potential comes from a fundamental action coupled to matter fields has remained unexplored. Moreover, the free parameters appearing in the metric lack, in their original formulation, a clear physical interpretation in terms of conserved charges. These shortcomings motivate the present investigation.
	
In this work, we propose a consistent matter source for the geometry introduced in Ref.~\cite{Khosravipoor:2023jsl}. Our key insight is that the deformation parameters can be naturally identified with a magnetic charge \(q_m\) through a dimensional analysis, thereby reducing the extended parameter space to a single physical scale associated with the magnetic sector. To achieve this, we employ a minimal extension of GR that incorporates a scalar field non-minimally coupled to NEDs via a coupling function \(W(\varphi)\). This framework allows us to reconstruct the scalar potential \(V(\varphi)\), the NED Lagrangian \(\mathcal{L}(F)\), and the coupling function \(W(\varphi)\) directly from the metric. The resulting model is both theoretically well founded and observationally testable, as we demonstrate through shadow constraints and thermodynamic analysis.

The paper is organized as follows: In Sec.~\ref{sec:um}, we present the space-time geometry and perform the reduction of the parameter space. In Sec.~\ref{sec:dois}, we detail the action framework, derive the field equations, and explicitly reconstruct the matter functions. The horizon structure and the black hole shadow are analyzed in Sec.~\ref{sec:rs}. Section~\ref{sec:termo} is dedicated to the study of black hole thermodynamics in the extended phase space. The topological analysis of the photon sphere and the thermodynamic parameter space is presented in Sec.~\ref{sec:topologia}. Our final considerations are presented in Sec.~\ref{sec:conclu}.
Throughout this work, we consider the metric signature to be \((+,-,-,-)\) and, unless otherwise stated, we use geometrized units where \(G = c = 1\).
	
\section{Spacetime geometry and parameter reduction \label{sec:um}}

We consider a static and spherically symmetric space-time described by the line element
\begin{equation}
	ds^2=A(r)dt^2-\frac{1}{A(r)}dr^2- r^2 (d\theta^2+\sin^2 \theta d \phi^2),\label{eq:ds}
\end{equation}
where the metric function $A(r)$ is given by
\begin{equation}
	A(r) = 1-\frac{2M}{r}+\frac{r^2}{l^2} +\alpha\,\frac{\beta^2+3r^2+3\beta r}{3r(\beta+r)^3}.
		\label{eq:metricaoriginal}
\end{equation}
Here, \(M\) denotes the black hole mass, $l=\sqrt{3/|\Lambda|}$ is the AdS radius with \(\Lambda\) the cosmological constant,  \(\alpha\) is a deformation parameter, and \(\beta\) is a constant with dimensions of length that controls the behavior of the effective energy density. 
This solution was originally proposed in Ref.~\cite{Khosravipoor:2023jsl} as an effective geometric space-time supported by a prescribed energy density. 

Our goal in this section is to identify a consistent matter source arising from a fundamental action. A closer inspection of the metric reveals that the parameter \(\alpha\) has dimensions of length squared, while \(\beta\) carries dimensions of length. This observation naturally suggests an interpretation in terms of a magnetically charged nonlinear electrodynamics (NED) source, for which the magnetic charge \(q_m\) provides the only relevant scale in the matter sector. 
In this context, dimensional consistency requires \(\alpha \propto q_m^2\) and \(\beta \propto q_m\).


Furthermore, in order to avoid pathological geometries and to ensure the existence of a well-defined event horizon, in accordance with the cosmic censorship conjecture, we restrict our attention to non-negative values of the parameters \(\alpha\) and \(\beta\). 
With this requirement we identify the free parameters of the metric as
\begin{equation}
	\alpha \rightarrow q_m^2, 
	\qquad 
	\beta \rightarrow |q_m|,
	\label{eq:alphaebeta}
\end{equation}
	
In what follows, we define \(\beta \equiv q\) and \(\alpha \equiv q^2\), where $q = |q_m|$. This choice reduces the original parameter space to a single physical parameter associated with the magnetic sector and ensures that the Schwarzschild--(A)dS limit is smoothly recovered in the limit \(q_m \to 0\). With these identifications, the space-time admits a consistent interpretation as being generated by a magnetic source, which will be explicitly constructed in the next section.

\section{Matter source }\label{sec:dois}

\subsection{Field equations and ansatz}

We consider the gravitational dynamics described by the action
\begin{eqnarray}
	S &=& \int d^4x \sqrt{-g}\bigl[
	R + 2\Lambda + 2\kappa^2\bigl(
	V(\varphi) - \varepsilon(\varphi)\nabla_a\varphi\nabla^a\varphi
		\nonumber \\
	&& + W(\varphi)\mathcal{L}(F)\bigr)\bigr],
		\label{eq:action}
\end{eqnarray}
where \(\varphi\) represents the scalar field and \(V(\varphi)\) its associated potential, \(\varepsilon (\varphi)\) is a function of the scalar field whose value defines the contribution of the latter as canonical or phantom, and \(W(\varphi)\) is a function that performs the non-minimal coupling between the scalar and electromagnetic fields, while both of them are minimally coupled to GR.

Varying the above action with respect to the metric, we obtain Einstein's equations in this case as
\begin{equation}
	G_{\mu \nu} - \Lambda g_{\mu \nu}= \kappa^2 \left( T^{(\varphi)}_{\mu \nu} + W(\varphi) T^{(\text{NED})}_{\mu \nu} \right),
	\label{eq:Einstein}
\end{equation}
where \(G_{\mu \nu}\) is the Einstein tensor defined by
\begin{equation}
	G_{\mu \nu} = R_{\mu \nu} - \frac{1}{2} R g_{\mu \nu},
\end{equation}
with \(R_{\mu \nu}\) the Ricci tensor and \(R\) the Ricci scalar. The right-hand side encodes the contributions from both the scalar field and the nonlinear electromagnetic sector.

The electromagnetic energy-momentum tensor is given by
\begin{equation}
	T^{(\text{NED})}_{\mu \nu} = g_{\mu \nu}\mathcal{L}(F) - \mathcal{L}_F F_{\mu}^{\ \alpha} F_{\alpha \nu},
	\label{7}	
\end{equation}
where \(\mathcal{L}_F \equiv \partial \mathcal{L}/\partial F\),
while the energy-momentum tensor associated with the scalar field reads
\begin{equation}
	\label{eq:tmunuphi}
	T^{(\varphi)}_{\mu \nu} = 2\varepsilon(\varphi) \nabla_{\mu} \varphi \nabla_{\nu} \varphi 
	- \varepsilon(\varphi) \nabla^{\alpha} \varphi \nabla_{\alpha} \varphi \, g_{\mu \nu} 
	+ V(\varphi) g_{\mu \nu}.
\end{equation}
These expressions show that both sectors contribute non-trivially to the space-time curvature, with the electromagnetic part being further modulated by the scalar-dependent coupling.

A key ingredient of this framework is the presence of the function \(W(\varphi)\), which must in general differ from the minimally coupled case \(W(\varphi)=1\). The action \eqref{eq:action} therefore represents a minimal extension of GR including a scalar degree of freedom non-minimally coupled to the electromagnetic sector. Such couplings arise naturally in scalar--tensor theories and are central to mechanisms of spontaneous or induced scalarization, in which the scalar field is triggered by the presence of matter fields rather than acting as a primary source of the geometry.

In this framework, the scalar field does not introduce new independent charges. Instead, its dynamics is entirely driven by the electromagnetic field through the coupling function \(W(\varphi)\), allowing the scalar sector to effectively reconstruct a prescribed geometry while maintaining a well-defined variational principle.

Now, varying the action with respect to the scalar field \( \varphi \), we obtain the corresponding equation of motion,
\begin{equation}
	2 \varepsilon(\varphi) \Box \varphi + \varepsilon'(\varphi) \nabla^a \varphi \nabla_a \varphi = - V'(\varphi) + W'(\varphi) \mathcal{L}(F),
	\label{eq:campoescalar}
\end{equation}
where \(\Box \equiv \nabla^\mu \nabla_\mu\) denotes the covariant d'Alembertian operator, and primes indicate derivatives with respect to the scalar field. This equation shows that the scalar dynamics are influenced not only by its potential \(V(\varphi)\), but also by its non-minimal coupling to the electromagnetic sector through the function \(W(\varphi)\).

Furthermore, varying the action with respect to the vector potential \(A_{\mu}\), we obtain the generalized electromagnetic field equations,
\begin{equation}
	\nabla_\mu \left( W(\varphi) \, \mathcal{L}_F \, F^{\mu\nu} \right) = 0.
	\label{eq:maxwell}
\end{equation}
These equations reduce to the standard Maxwell equations in the appropriate limit, but here they explicitly incorporate the coupling to the scalar field via \(W(\varphi)\), modifying the effective electromagnetic dynamics.

It is important to note that, in this framework, the scalar field equations \eqref{eq:campoescalar} and the electromagnetic equations \eqref{eq:maxwell} are coupled. From the Einstein equations \eqref{eq:Einstein} we find
\begin{equation}
	\frac{1-r A'-A-\Lambda r^2}{r^2}
	-\kappa^2\left(A\,\varepsilon\,\varphi'^2+\mathcal{L}\,W
	+V\right) = 0,
		\label{eq:00}
\end{equation}
\begin{equation}
	\frac{1}{r^2}-\frac{r A'+A}{r^2}
	+\kappa^2 A\,\varepsilon\,\varphi'^2 -\Lambda
		-\kappa^2 (\mathcal{L}\,W+V)=0,
		\label{eq:11}
\end{equation}
\begin{equation}
	\Lambda + \frac{A''}{2} + \frac{A'}{r}  +\kappa^2\Big[
		A\varepsilon\varphi'^2 +W\!\left(\mathcal{L}
		-\frac{q^2\mathcal{L}_F}{r^4}\right) +V \Big] =0.
		\label{eq:22}
\end{equation}

We now solve Eqs.~\eqref{eq:00} and \eqref{eq:22} to determine the electromagnetic Lagrangian density \(\mathcal{L}\) and its derivative \(\mathcal{L}_F\), obtaining
\begin{equation}
	\mathcal{L} = \frac{
		1 - r\left(A' + \kappa^2 r V\right)
		- A\left(\kappa^2 r^2 \varepsilon\,\varphi'^2 + 1\right)
		+ \Lambda r^2 }{\kappa^2 r^2 W},
	\label{eq:L}
\end{equation}
and
\begin{equation}
	\mathcal{L}_F = \frac{r^2 \left(r^2 A'' - 2 A + 2\right)}{2 \kappa^2 q^2 W},
	\label{eq:LF}
\end{equation}
respectively. These expressions follow directly from Einstein's equations and encode the effective NED compatible with the assumed geometry.

Since \(\mathcal{L}\) and \(\mathcal{L}_F\) are obtained independently, consistency with an underlying electromagnetic theory requires that they satisfy the relation
\begin{equation}
	\mathcal{L}_F - \frac{\partial \mathcal{L}}{\partial r} \left( \frac{\partial F}{\partial r} \right)^{-1} = 0,
\end{equation}
which ensures that \(\mathcal{L}_F = \partial \mathcal{L}/\partial F\) holds identically. This condition acts as a non-trivial constraint selecting physically admissible solutions.

Substituting the above expressions back into the remaining Einstein equations, one finds that all differential equations reduce to a single algebraic condition involving the scalar sector,
\begin{equation}
	2 \kappa^2 A(r)\,\varepsilon(r)\,\varphi'(r)^2 = 0,
	\label{eq:epsolon}
\end{equation}
which severely restricts the possible configurations of the scalar field and its coupling.

Since the metric function \(A(r)\) is not identically zero in the space-time region of interest, Eq.~\eqref{eq:epsolon} leads to two possible branches: either the scalar field is constant, \(\varphi'(r)=0\), or the kinetic coupling function vanishes, \(\varepsilon(r)=0\). The first case corresponds to a trivial scalar configuration, which does not play an active role in the reconstruction of the matter sector and will therefore not be pursued. We thus focus on the second branch, \(\varepsilon(r)=0\), for which the scalar field remains nontrivial and continues to interact with the electromagnetic sector through the coupling function \(W(\varphi)\).

In this branch, the scalar field profile can be chosen freely, providing additional flexibility in constructing solutions. The scalar potential consistent with the geometry and the chosen coupling function \(W(\varphi)\) is then determined directly from the scalar field equation \eqref{eq:campoescalar}, yielding
\begin{eqnarray}
	V(r) = \frac{W}{\kappa^2}\left(\int \frac{W' \left(r A' + A + \Lambda r^2 - 1\right)}{r^2 W^2} \, dr \right).
	\label{eq:pot}
\end{eqnarray}
This expression shows that the potential is not arbitrary, but is fixed by the interplay between the geometry and the scalar--electromagnetic coupling.

\subsection{Reconstruction of the matter sector}

With all relevant physical quantities expressed in terms of the metric function \(A(r)\), we can now explicitly solve the system. Substituting Eq.~\eqref{eq:alphaebeta} into \eqref{eq:metricaoriginal}, we obtain
\begin{equation}
	\begin{aligned}
		A(r) = 1 - \frac{2 M}{r} + \frac{q^2 \left(q^2 + 3 q r + 3 r^2\right)}{3 r (q+r)^3} - \frac{\Lambda r^2}{3},
	\end{aligned}
	\label{eq:metrica}
\end{equation}
which describes the space-time geometry in terms of the mass \(M\), cosmological constant \(\Lambda\), and magnetic charge \(q\). Since the matter sector introduces a single magnetic charge, we restrict the parameter space by identifying the integration constants accordingly, thereby avoiding redundant degrees of freedom.

Substituting Eq.~\eqref{eq:metrica} into Eq.~\eqref{eq:LF}, we find
\begin{equation}
	\mathcal{L}_F \equiv \frac{d \mathcal{L}}{d F} = \frac{2 r^5}{\kappa^2 W (q+r)^5},
	\label{eq:principal}
\end{equation}
where the electromagnetic invariant is given by \(F = q^2/2r^4\). To proceed, we impose the condition
\begin{equation} \label{eq:condf}
	\frac{2 r^5}{\kappa^2 W (q+r)^5} = F(r)^n,
\end{equation}
with \(n\) a natural number. This ansatz allows one to fix the form of the coupling function \(W(\varphi)\) consistently with the nonlinear electrodynamics sector.

Furthermore, with this choice we obtain the explicit form of the coupling function $W(r)$ as
\begin{equation}
	W(r) =  \frac{2^{n+1} r^5 \left(\frac{q^2}{r^4}\right)^{-n}}{\kappa ^2 (q+r)^5}.
	\label{eq:wmr}
\end{equation}
This expression completely specifies how the non-linear electromagnetic sector couples to the geometry through the radial coordinate. Substituting $W(r)$ into the field equations, we can determine both the Lagrangian density and the corresponding potential. In particular, the Lagrangian density $\mathcal{L}(r)$ is given by
\begin{equation}
	\mathcal{L}(r) =  \frac{2^{-n-1} q^{2 n+2} r^{-4 (n+1)}}{n+1},
	\label{eq:Lrm}
\end{equation}
while the associated potential takes the form
\begin{equation}
	V(r) = \frac{q^2 (n q+n r+q)}{\kappa ^2 (n+1) (q+r)^5}.
	\label{eq:V(r)}
\end{equation}

In order to rewrite the Lagrangian density in terms of the field strength scalar $F$, we use its dependence on the radial coordinate (valid for $q>0$), namely,
\begin{equation}
	r=  \left(\frac{q^2}{2F }\right)^{1/4}.
\end{equation}
Substituting this relation into $\mathcal{L}(r)$ and simplifying, we can express the theory entirely in terms of $F$. This yields
\begin{equation}
	\mathcal{L}(F) = \frac{F^{n+1}}{n+1},
	\label{eq:Lmagnetico}
\end{equation}
which defines a one-parameter family of electromagnetic Lagrangians, characterized by the integer $n \ge 0$\footnote{Models of this kind have been considered previously in the literature, see e.g. \cite{Hassaine:2007py}.}. The case $n=0$ corresponds to standard linear electrodynamics, for which $\mathcal{L}(F)=F$, while values $n>0$ describe NEDs, with higher-order contributions in the electromagnetic invariant $F$.

NED models, including the family above, are well known to generate effective matter sources with non-trivial internal structure and play a central role in the construction of regular black hole geometries \cite{Ansoldi:2008jw}. Therefore, the present framework naturally accommodates both linear and non-linear electromagnetic sources. In this sense, the parameter $n$ controls the degree of non-linearity of the electromagnetic field and determines the effective matter content supporting the geometry.

In the following, we shall keep $n$ arbitrary whenever possible, thus allowing for a general NED interpretation of the source. At this stage, we exploit the freedom in the choice of the scalar field profile and adopt
\begin{equation}
	\varphi(r) = \frac{1}{ \kappa }\,\tan^{-1}\left(\frac{r}{q}\right),
		\label{eq:phi(r)}
\end{equation}
which is a functional form commonly used in the literature to describe the scalar field~\cite{Rodrigues:2023vtm,Alencar:2024yvh}.
This function tends to zero when $r \rightarrow 0$ and to $\pi/(2\kappa)$ when $r \rightarrow \infty$, and is therefore bounded between these values. We can invert Eq.~\eqref{eq:phi(r)} and then substitute the result into Eqs. ~\eqref{eq:V(r)} and~\eqref{eq:wmr} to write
\begin{equation}
	W( \varphi) =\frac{2^{n+1} \left(\frac{\cot ^4(\kappa  \varphi)}{q^2}\right)^{-n}}{\kappa ^2 (\cot (\kappa 
	\varphi)+1)^5},
		\label{eq:wmphi}
\end{equation}
and
\begin{equation}
	V( \varphi) =\frac{n \tan (\kappa  \varphi)+n+1}{\kappa ^2(n+1) q^2 (\tan (\kappa  \varphi)+1)^5},
		\label{eq:Vmphi}
\end{equation}
for the coupling function and the scalar potential, respectively.

\begin{figure}[t!]
	\centering
	\includegraphics[width=\columnwidth]{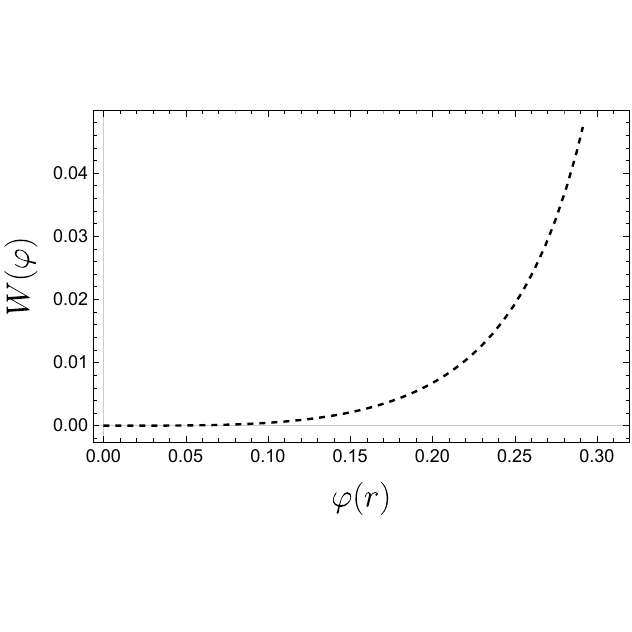}
	\caption{The coupling function $W(\varphi)$ given by Eq.~\eqref{eq:wmphi}, for $M=1$, $q=0.5$, and $n=0$.}
	\label{fig:wm(phi)1}
\end{figure}

\begin{figure}[t!]
	\centering
	\includegraphics[width=\columnwidth]{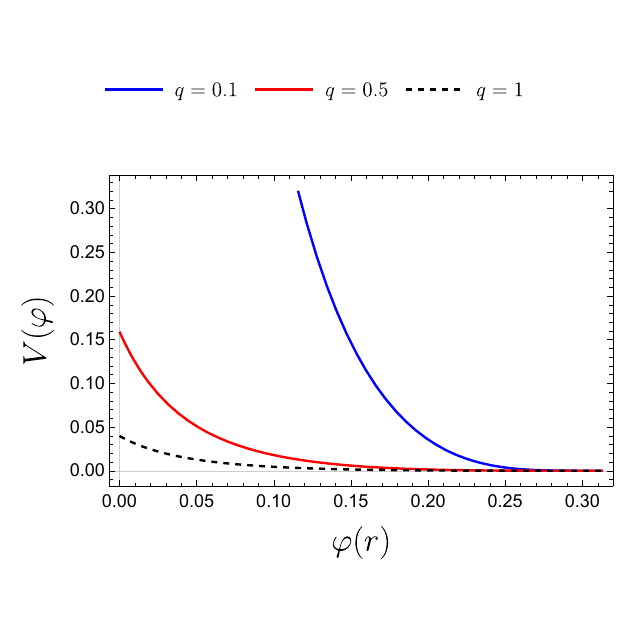}
	\caption{The potential $V(\varphi)$ given by Eq.~\eqref{eq:Vmphi}, for $M=1$, and $n=0$.}
		\label{fig:Vm(phi)1}
\end{figure}

The coupling function $W(\varphi)$ and the scalar potential $V(\varphi)$ are shown in Figs.~\ref{fig:wm(phi)1} and~\ref{fig:Vm(phi)1}, respectively, for the linear electrodynamics case $n=0$. The function $W(\varphi)$ is positive and regular throughout the entire domain of the scalar field, ensuring a well-defined non-minimal coupling between the scalar and electromagnetic sectors. It is clear from Eq.~\eqref{eq:wmphi} that for $n=0$ this function does not depend on $q$. The scalar potential $V(\varphi)$ is also regular and bounded, and increasing the charge decreases its values. This behavior is physically appealing because it implies that stronger magnetic fields suppress the scalar potential, thereby naturally taming the backreaction on the geometry.

\section{Horizon structure and black hole shadow}

\subsection{Horizons and critical charge}

In order to normalize the horizon structure of our black holes, let us first normalize all parameters by the mass via the variable substitutions $\rho = r/M$, $Q = q/M $, and $\lambda = \Lambda M^2$ in Eq.~\eqref{eq:metrica}, we obtain
\begin{equation}
	A(\rho) = 1-\frac{\lambda  \rho ^2}{3}-\frac{2}{\rho }+\frac{Q^2 \left(3 \rho ^2+Q^2+3 \rho  Q\right)}{3 \rho  (\rho
		+Q)^3},
	\label{eq:metrica2}
\end{equation}

The locations of the horizons are determined by the real, positive roots of $A (\rho) = 0$. For suitable ranges of the magnetic charge 
$Q$, the spacetime admits both an event horizon and a cosmological horizon. The critical value $Q_c$ corresponds to an extremal configuration in which the two horizons (event and inner) merge and is obtained by simultaneously solving
\begin{equation}
	A (\rho) = 0, \qquad \frac{\partial A (\rho)}{\partial \rho} = 0.
	\label{eq:conh}
\end{equation}

\begin{figure}[t!]
	\centering
	\includegraphics[width=\columnwidth]{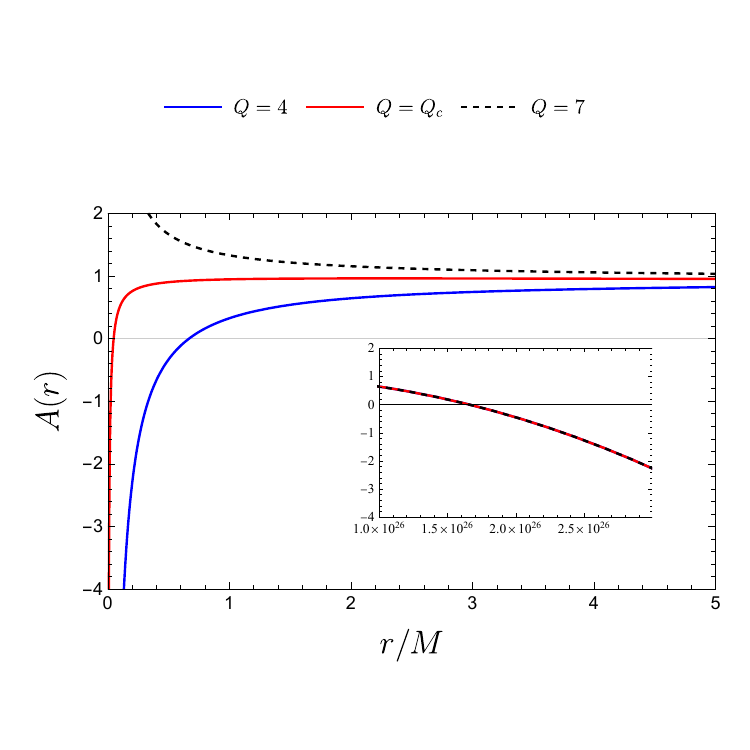}
	\caption{ Metric function $A(\rho)$ given in Eq.~\eqref{eq:metrica2} with $\Lambda=  1.08947 \times 10^{-52}$ and $n=0$.}
		\label{fig:A(r)}
\end{figure}

This analysis is essential to restrict the parameter space to configurations describing black holes. In the following, we therefore consider only values $Q<Q_c$, for which the solution exhibits a well-defined event horizon. Fig.~\ref{fig:A(r)} illustrates the behavior of $A (\rho)$ for representative values of the charge. We solve Eqs.~\eqref{eq:conh} numerically and find $Q_c = 5.85$.

For $Q < Q_c$ (blue curve), the metric function possesses two distinct zeros, corresponding to the event and cosmological horizons. At the critical value $Q = Q_c$ (red curve), the two horizons merge, resulting in an extremal black hole with zero surface gravity. For $Q > Q_c$ (dashed black curve), no event horizon is present, and the spacetime describes a horizonless compact object rather than a black hole. 

It is worth noting that the behavior of $A(\rho)$ near the origin changes qualitatively across the critical charge. As $\rho \rightarrow 0$, the divergence of the metric function is negative for $Q<Q_c$ and positive for $Q>Q_c$, signaling a modification of the causal structure in the supercritical regime. This transition is reminiscent of the behavior found in Reissner--Nordström black holes, where overcharging destroys the horizon. Thus, the critical charge marks the boundary between black hole and horizonless configurations, a feature that will be important when comparing with observational constraints.

\subsection{Shadow radius \label{sec:rs}}
	
The shadow of a black hole corresponds to a dark region in the observer's sky associated with photon trajectories that are captured by the gravitational field of the black hole. Its size is determined by the projection of the surface of unstable bound circular photon orbits on the observer's screen, and provides a direct observational probe of the underlying space-time geometry.
	
For a static and spherically symmetric space-time of the form \eqref{eq:ds}, the shadow radius measured by a distant observer located at radial coordinate $r_0$ is given by \cite{Perlick:2021aok}
\begin{equation}
	r_s = r_p \sqrt{\frac{A(r_0)}{A(r_p)}} ,
		\label{eq:r_shadow}
\end{equation}
where $r_p$ denotes the radius of the photon sphere. 
	
The photon sphere radius is determined by the condition \cite{Claudel:2000yi}
\begin{equation}
	\frac{2}{r_p} = \frac{A'(r_p)}{A(r_p)} ,
		\label{eq:fotonesfera}
\end{equation}
Substituting the dimensionless metric function~\eqref{eq:metrica2} into Eq.~\eqref{eq:fotonesfera}, we obtain the algebraic equation
\begin{equation}
	\begin{aligned}
		&2 (\rho -3) \rho ^4 + Q^5 + 6 (\rho -1) Q^4 \\
		&+ 2 \rho (7 \rho -12) Q^3 + 4 \rho ^2 (4 \rho -9) Q^2 \\
		&+ 8 (\rho -3) \rho ^3 Q = 0 ,
	\end{aligned}
		\label{eq:fotonN}
\end{equation}
which is solved numerically to determine the photon sphere radius $\rho_p = r_p/M$.
	
To compare the model with observations, we use the predicted shadow radius and the Event Horizon Telescope measurements of Sagittarius A*. The mass, $(3.951 \pm 0.047) \times 10^6 M_{\odot}$, and distance, $7.953 \pm 0.050 \pm 0.032$ kpc, used in our analysis are estimates provided by the Keck collaboration \cite{Do:2019txf}. Following Refs.~\cite{EventHorizonTelescope:2022xqj,Vagnozzi:2022moj}, the deviation of the shadow radius from the Schwarzschild value $r_{s,\mathrm{Sch}} = 3\sqrt{3}\,M$ is parametrized as
\begin{equation}
	\frac{r_s}{M} = (1+\delta)\,3\sqrt{3} ,
\end{equation}
which leads to the observational bounds
\begin{equation}
	4.55 \lesssim \frac{r_s}{M} \lesssim 5.22 \qquad (1\sigma),
\end{equation}
and
\begin{equation}
	4.21 \lesssim \frac{r_s}{M} \lesssim 5.56 \qquad (2\sigma).
\end{equation}
	
\begin{figure}[t!]
    \centering
    \includegraphics[width=\columnwidth]{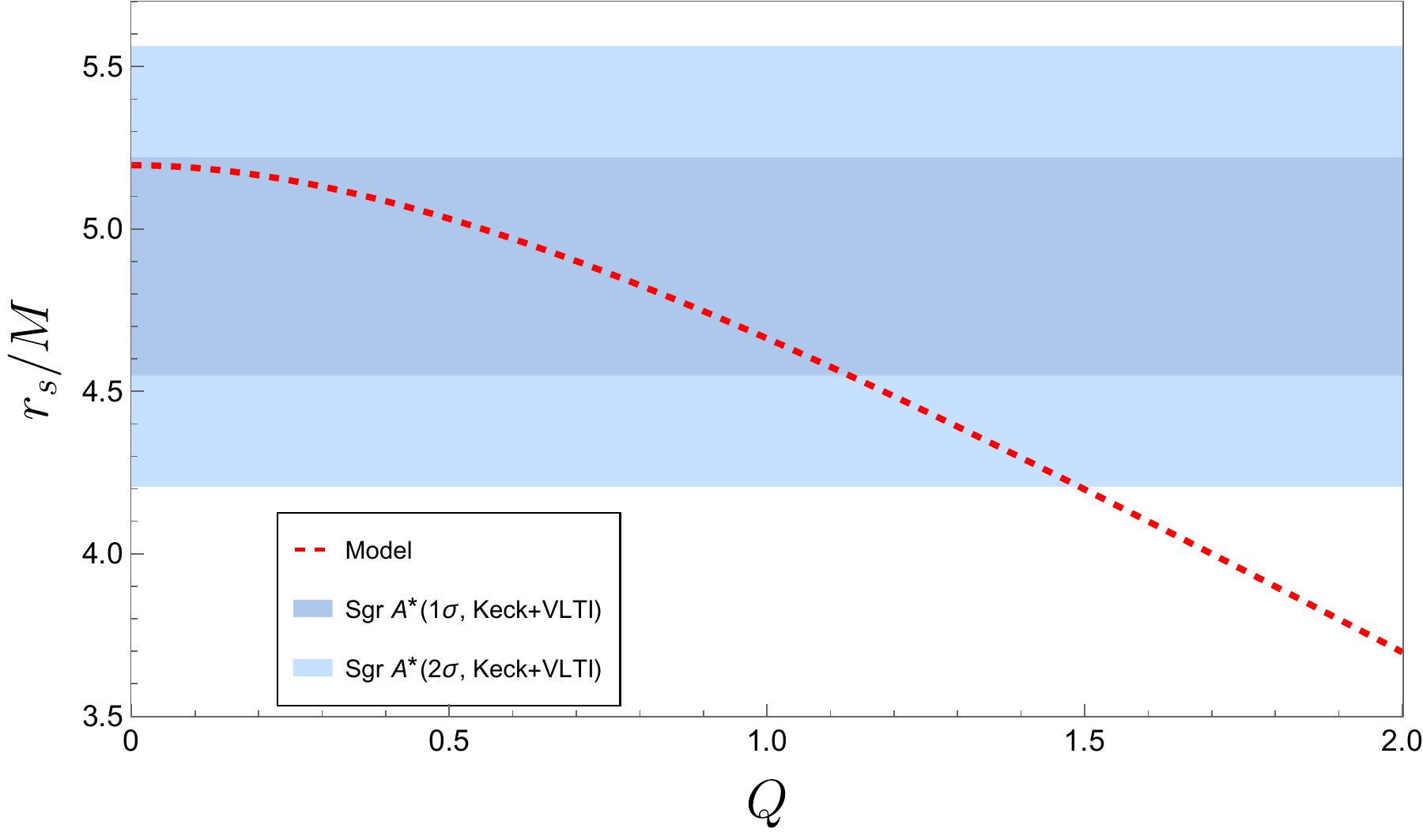}
    \caption{Black hole shadow radius $r_s/M$ as a function of the magnetic charge per unit mass $Q$, for and observer located at $r_0 \sim 8\  \text{kpc} $ and with $\Lambda=  1.08947 \times 10^{-52}$ . The horizontal shaded regions represent the $1\sigma$ and $2\sigma$ observational bounds from the Event Horizon Telescope for Sagittarius~A*. 
    Values of $Q$ within these regions are compatible with current EHT data.}
    \label{fig:sombra}
\end{figure}

Fig.~\ref{fig:sombra} displays the shadow radius for representative values of
the magnetic charge $Q$, highlighting the regions of parameter space compatible
with the EHT constraints. We can see that the shadow radius stays within $(1\sigma)$ for $Q < 1.12$, and within $(2\sigma)$ until $Q < 1.48$. It is worth noting that, although we have considered $r_s$ as given by Eq. \eqref{eq:r_shadow} under the assumption that the metric is not asymptotically flat, the practical result for the value of $\Lambda$ under consideration is $A(r_0) \approx 1$ to all practical purposes.  On \cite{Perlick:2021aok} the authors suggest that this indicates that, in a space time where the cosmological constant accounts for the cosmic acceleration, there is no detectable difference on the black hole shadow.

\section{Black hole thermodynamics \label{sec:termo}}

In this section, we analyze the thermodynamic properties of the black hole solution described in the previous sections. Our goal is to explore how the NED and scalar field configurations influence the standard thermodynamic quantities. We adopt the framework of extended black hole thermodynamics, in which the cosmological constant is treated as a dynamical quantity and is interpreted as a thermodynamic pressure according to
\begin{equation}
	P = -\frac{\Lambda}{8\pi}.
\end{equation}
Within this formalism, the black hole mass is naturally identified with the enthalpy of the space-time, rather than its internal energy, which allows us to consistently define thermodynamic conjugate variables and study phase transitions in analogy with standard thermodynamic systems.

\subsection{Mass as a thermodynamic potential}

The event horizon radius $r_h$ is determined by the largest real root of the equation $A(r_h)=0$, where $A(r)$ is the metric function given in Eq.~\eqref{eq:metrica}. This root represents the outermost surface from which no light can escape, and therefore characterizes the size of the black hole. Solving this equation for the mass parameter $M$, we obtain
\begin{equation}
	M = M(r_h,q,P),
\end{equation}
which expresses the mass as a function of the horizon radius, the magnetic charge, and the thermodynamic pressure.

Using the Bekenstein--Hawking area law, the black hole entropy is given by
\begin{equation}
	S = \frac{\mathcal{A}}{4} = \pi r_h^2,
\end{equation}
where $\mathcal{A}$ is the area of the event horizon. This relation allows us to express the horizon radius in terms of the entropy as
\begin{equation}
	r_h = \sqrt{\frac{S}{\pi}}.
\end{equation}
Substituting this relation and the identification $\Lambda=-8\pi P$ into the expression for the mass, we obtain the enthalpy of the black hole as a function of the entropy $S$, pressure $P$, and magnetic charge $q$:
\begin{eqnarray}
	&&M(S,P,q) = \frac{1}{6\sqrt{\pi}\,(\sqrt{\pi}q+\sqrt{S})^3} 
	\nonumber \\
	&&\qquad \times \Big[\pi^2 q^4 + 2\pi^{3/2} q^3 \sqrt{S}\,(3+4PS) 
	\nonumber \\
	&& \quad + 12\pi q^2 S\,(1+2PS) 
	\nonumber \\
	&& \quad + 3\sqrt{\pi} q S^{3/2}\,(3+8PS) 
	+ S^2(3+8PS)
	\Big].
	\label{eq:M}
\end{eqnarray}

Eq.~\eqref{eq:M} serves as the fundamental thermodynamic potential of the system, analogous to the enthalpy in standard thermodynamics. From this function, all other thermodynamic quantities—such as temperature, pressure, and specific heats—can be derived systematically, providing a complete description of the black hole’s thermodynamic behavior.

\subsection{First law and thermodynamic quantities}
	
The temperature of the black hole, the thermodynamic volume, and the magnetic potential conjugate to the charge are obtained from the enthalpy via
\begin{equation}
	T = \left(\frac{\partial M}{\partial S}\right)_{P,q}, \ \  V = \left(\frac{\partial M}{\partial P}\right)_{S,q} , \ \  A_q = \left(\frac{\partial M}{\partial q}\right)_{S,P}.
\end{equation}
Thus, we find directly from Eq.~\eqref{eq:M} 
\begin{equation}
	T =  \frac{8 P S-\frac{\pi  q^2 S}{\left(\sqrt{\pi } q+\sqrt{S}\right)^4}+1}{4 \sqrt{\pi } \sqrt{S}},
		\label{eq:T}
\end{equation}
\begin{equation}
	V =  \frac{4 S^{3/2}}{3 \sqrt{\pi }},
\end{equation}
\begin{equation}
	A_q =  \frac{\pi ^2 q^4+4 \pi ^{3/2} q^3 \sqrt{S}+6 \pi  q^2 S+6 \sqrt{\pi } q S^{3/2}}{6 \left(\sqrt{\pi } q+\sqrt{S}\right)^4}. 	
\end{equation}

We note that for positive pressure ($P>0$), corresponding to an AdS background, the temperature remains strictly positive for all physically allowed values of the entropy. In particular, the temperature diverges in the small entropy limit, indicating the breakdown of the thermodynamic description for very small black holes. This is a typical feature of AdS black holes and is related to the fact that the semiclassical approximation fails when the horizon radius approaches the Planck scale.
	
In the extended phase space, the mass $M$ is interpreted as the enthalpy and is expressed as a function of the entropy $S$, the pressure $P$, and the magnetic charge $q$, i.e., $M=M(S,P,q)$. Under the scaling transformation $r_h \rightarrow \lambda r_h$, the thermodynamic variables scale as $S \sim r_h^2$, $P \sim r_h^{-2}$, and $q \sim r_h$, implying that the mass is a homogeneous function of degree $\delta=1/2$. As a consequence of Euler's theorem for homogeneous functions,
the Smarr relation takes the form
\begin{equation}
	M = 2(TS - PV) + A_q q.
\end{equation}
The first law of black hole thermodynamics is given by
\begin{equation}
	dM = T\,dS + A_q\,dq + V\,dP,
\end{equation}
which is explicitly verified by the thermodynamic potentials obtained from the enthalpy $M(S,P,q)$.
	
\begin{figure}[t!]
	\centering
	\includegraphics[width=\columnwidth]{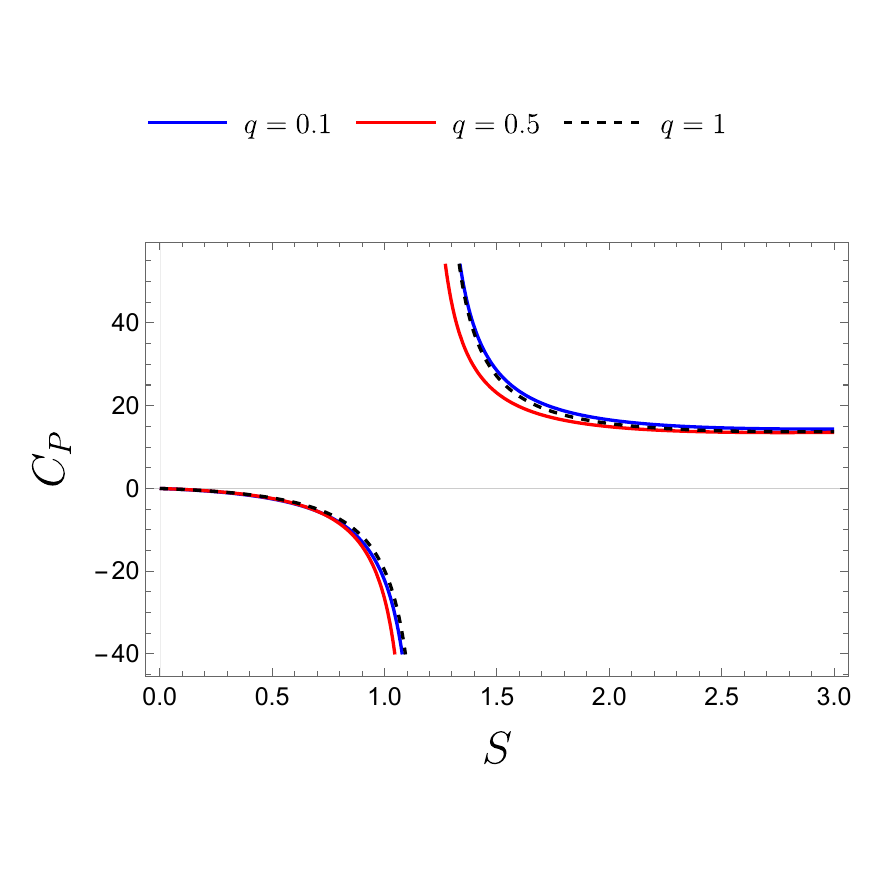}
	\caption{The heat capacity given by Eq.~\eqref{eq:Cp} with $P=0.1$.}
		\label{fig:cp}
\end{figure}

We now focus on the thermodynamic stability of the black hole.~First, we investigate the local stability through the behavior of the heat capacity at constant pressure. 
Configurations with $C_P>0$ are locally stable, while negative values of $C_P$ signal thermodynamic instability. The divergence of $C_P$ indicates a Davies-type second-order phase transition separating different black hole branches. For this solution it is
\begin{equation}
	C_P = T\left(\frac{\partial S}{\partial T}\right)_P
	= \frac{T}{\left(\frac{\partial T}{\partial S}\right)_P}.
\end{equation}
Using the explicit expression for the Hawking temperature $T(S,P,q)$, we obtain
\begin{equation}
		C_P = \frac{2 S \left(1 + 8 P S - \dfrac{\pi q^2 S}{(\sqrt{\pi} q + \sqrt{S})^4}\right)
		}{8 P S - \dfrac{\pi q^2 S(\sqrt{\pi} q - 3\sqrt{S})}{(\sqrt{\pi} q + \sqrt{S})^5}-1}.
		\label{eq:Cp}
\end{equation}
Fig.~\ref{fig:cp} shows the behavior of $C_P$ as a function of the entropy for three different values of the charge parameter. For small entropies, the heat capacity is negative, indicating an unstable branch, whereas for larger entropies it becomes positive, signaling a stable large black hole phase. The transition occurs at a single point where $C_P$ diverges, which marks a second-order phase transition between the two branches.

To analyze the global thermodynamic stability, we study the Gibbs free energy defined as
\begin{equation}
	G = M - T S,
\end{equation}
where the mass $M$ is interpreted as the enthalpy in the extended phase space. The Hawking--Page phase transition is characterized by a change of sign of the Gibbs free energy, $G=0$, separating thermal AdS from the black hole phase. This transition is of first order and is not associated with a divergence of the heat capacity. While the divergence of the heat capacity signals a second-order (Davies-type) phase transition between different black hole branches, the Hawking--Page transition describes a first-order transition governed by the global behavior of the Gibbs free energy. Directly from \eqref{eq:M} and \eqref{eq:T} we obtain
\begin{eqnarray}
	G&=&\frac{1}{12\sqrt{\pi}\left(\sqrt{\pi}q+\sqrt{S}\right)^4}
		\Big[2\pi^{5/2}q^5+\pi^2 q^4 \sqrt{S}(3-8PS)  
			\nonumber \\
	&&+8\pi^{3/2} q^3 S(3-4PS)+3\pi q^2 S^{3/2}(9-16PS) 
		\nonumber \\
	&&+4\sqrt{\pi} q S^2(3-8PS)+S^{5/2}(3-8PS)\Big].
		\label{eq:G(S)}
\end{eqnarray}

\begin{figure}[t!]
	\centering
	\includegraphics[width=\columnwidth]{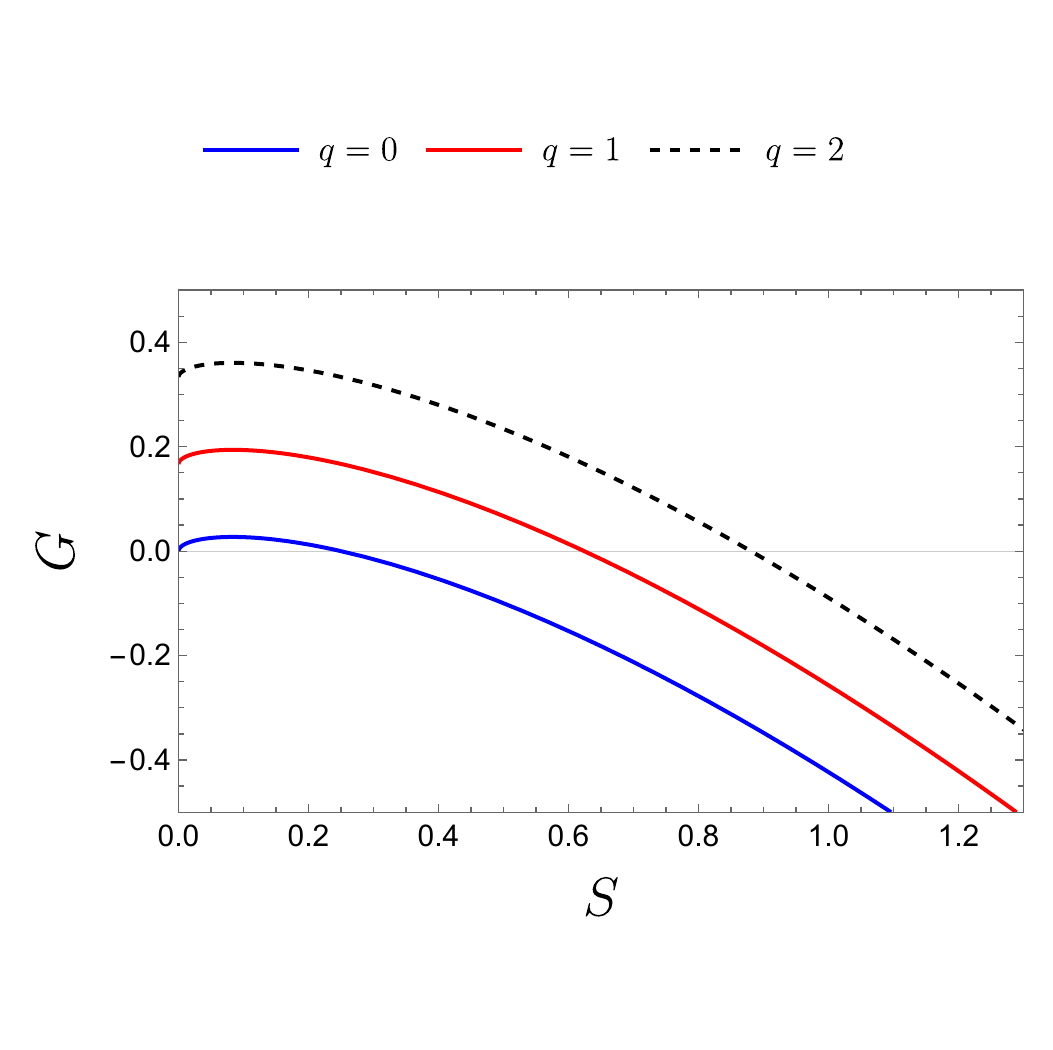}
	\caption{The Gibbs free energy given by Eq.~\eqref{eq:G(S)} with $P=1.5$. We can see that the increase of the charge leads to stable solutions represented by less negative values of $G$.}
		\label{fig:G}
\end{figure}
\begin{figure}[t!]
	\centering
	\includegraphics[width=\columnwidth]{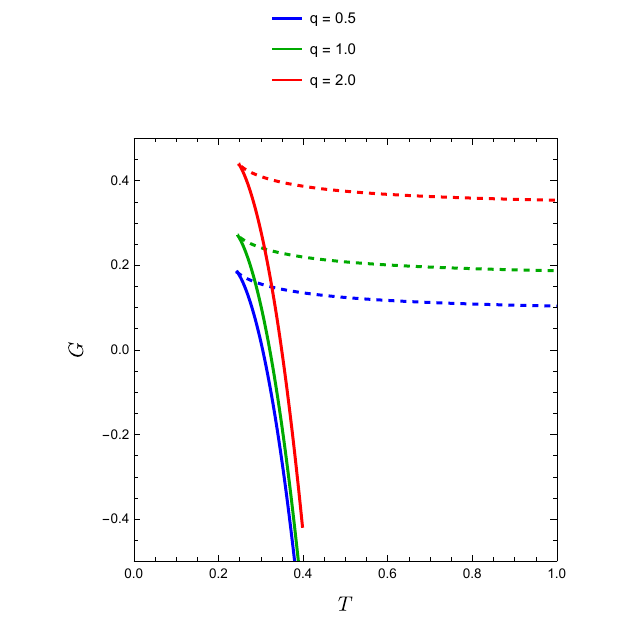}
	\caption{Gibbs free energy $G$ as a function of the Hawking temperature $T$ for fixed pressure $P$ and different values of the charge parameter $q$. The curves are parametrized by the entropy $S$. Dashed lines correspond to the small black hole branch, while solid lines represent the large black hole branch. The intersection of each curve with the horizontal axis ($G=0$) signals the Hawking--Page phase transition.
	For a given pressure, increasing the charge parameter shifts the phase transition to higher temperatures and modifies the stability structure of the black hole configurations.}
		\label{fig:GXT}
\end{figure}

Figs.~\ref{fig:G} and \ref{fig:GXT} summarize the global thermodynamic behavior of the black hole. In Fig.~\ref{fig:G}, the sign of the Gibbs free energy determines the preferred thermodynamic phase, with negative values indicating globally stable black hole configurations. Fig.~\ref{fig:GXT} provides a clearer picture of the phase structure by displaying the Gibbs free energy as a function of the Hawking temperature. The intersection with the horizontal axis corresponds to the Hawking--Page phase transition, separating the thermal AdS phase from the black hole phase. As the charge parameter increases, the transition temperature shifts to higher values, while the qualitative structure of the phase diagram remains unchanged. This behavior is qualitatively similar to that of Reissner--Nordström--AdS black holes, reinforcing the consistency of our solution.

\section{Topology}\label{sec:topologia}

\subsection{Topology of the photon sphere}
	
The photon sphere in static and spherically symmetric space-times
is determined by the condition \eqref{eq:fotonesfera}. Following the topological construction of Ref.~\cite{Wei:2020rbh}, it can be characterized as a zero of a suitable vector field defined in the $(r,\theta)$ plane. For black hole space-times possessing an event horizon, the total topological charge computed outside the horizon is
\begin{equation}
	\mathcal{Q}_{\text{tot}}=-1,
\end{equation}
whereas space-times without a horizon yield $\mathcal{Q}_{\text{tot}}=0$. Thus, the existence of a standard unstable photon sphere is topologically protected in black hole geometries. In order to implement such a topological construction, let us introduce an everywhere regular auxiliary potential function defined by
\begin{equation}
	H(r,\theta) = \frac{\sqrt{A(r)}}{r\,\sin\theta}.
		\label{eq:H}
\end{equation}
From this potential, one constructs a vector field $\boldsymbol{\Phi}=(\Phi_r,\Phi_\theta)$ on the $(r,\theta)$ plane, with components
\begin{equation}
	\Phi_r = \sqrt{A(r)}\,\partial_r H,
	\qquad
	\Phi_\theta = \frac{1}{r}\,\partial_\theta H .
\end{equation}
For the metric \eqref{eq:metrica}, these components take the explicit form
\begin{align}
	\Phi_r(r,\theta) &=
	\frac{\csc\theta}{2r^2\sqrt{A(r)}}
	\left[r A'(r) - 2A(r)\right], \label{eq:Phir} \\
	\Phi_\theta(r,\theta) &=
	-\frac{\cot\theta\,\csc\theta\,\sqrt{A(r)}}{r}.
		\label{eq:Phitheta}
\end{align}
	
The associated normalized vector field is defined as
\begin{equation}
	\mathbf{n} = \frac{\boldsymbol{\Phi}}
	{\sqrt{\Phi_r^2+\Phi_\theta^2}}
	= (n_r,n_\theta),
		\label{eq:nvector}
\end{equation}
which is regular everywhere except at the zeros of $\boldsymbol{\Phi}$.
	
It is important to note that the radial component $\Phi_r$ vanishes when the following condition holds:
\begin{equation}
	r A'(r) - 2A(r) = 0,
		\label{eq:photon_condition}
\end{equation}
which coincides with the standard condition for the existence of a photon sphere in static and spherically symmetric space-times. Hence, the photon sphere corresponds to a defect (zero) of the vector field $\boldsymbol{\Phi}$, allowing its characterization through topological invariants.
	
To investigate the topological properties of the photon sphere, we now specialize the general formalism to the black hole solution discussed in this work. We consider a static and spherically symmetric spacetime described by the line element \eqref{eq:ds} where $A(r)$ is the metric function given in Eq.~\eqref{eq:metrica}.
In order to explicitly visualize the topological structure of the
photon sphere, we fix representative values of the parameters.
Without loss of generality, we set
\begin{equation}
	M=1, \qquad q=M, \qquad \Lambda=-0.1\,M^{-2},
\end{equation}
which correspond to an asymptotically AdS space-time. For this choice, the metric function admits a single event horizon, located at
\begin{equation}
	r_h \simeq 1.61,
\end{equation}
while the photon sphere is found at
\begin{equation}
	r_{\rm ps} \simeq 2.64,
\end{equation}
lying outside the event horizon and therefore representing a physically accessible surface
	
\begin{figure}[t!]
	\centering
	\includegraphics[width=\columnwidth]{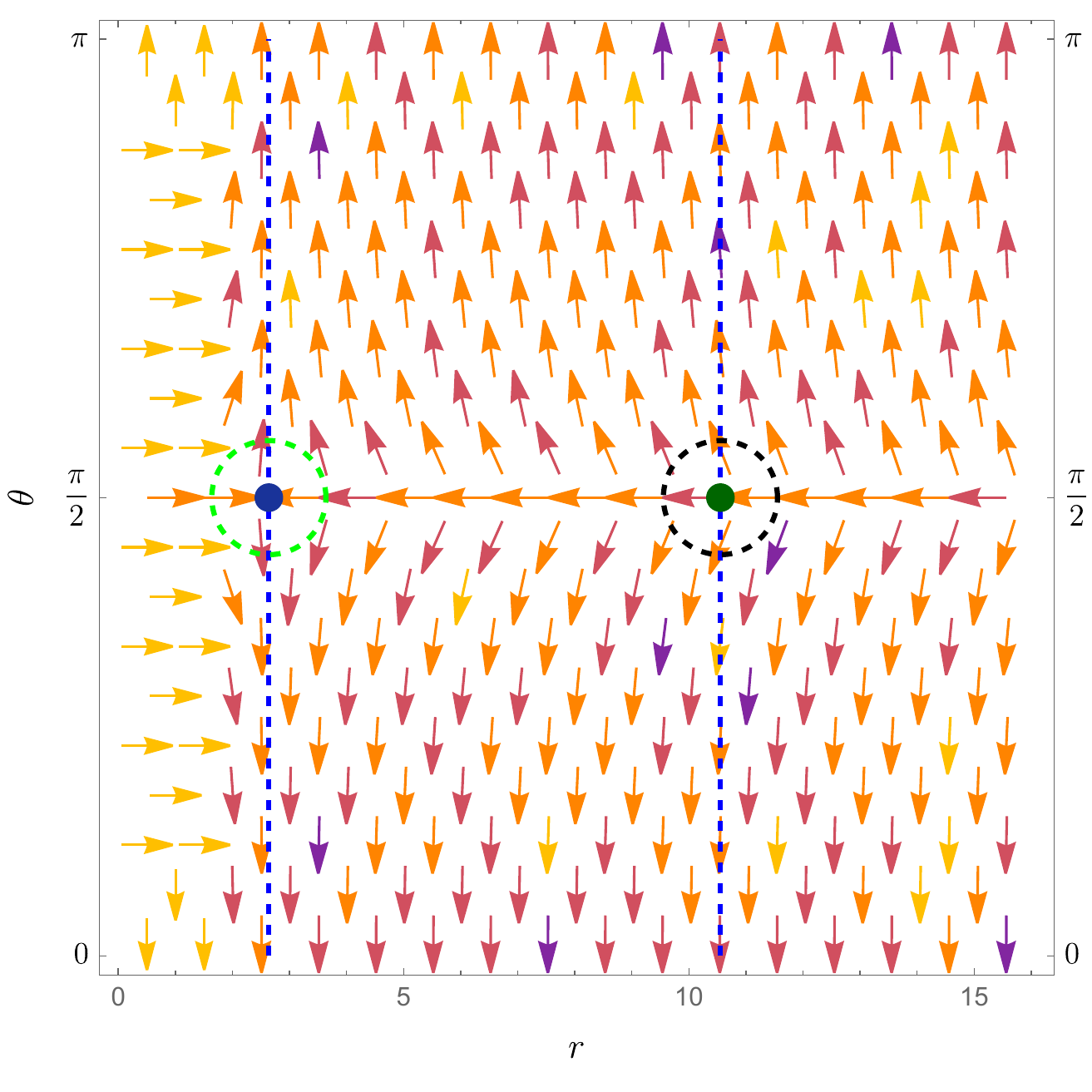}
	\caption{Normalized vector field $\mathbf{n}=(n_r,n_\theta)$ defined in Eq.~\eqref{eq:nvector} on the $(r,\theta)$ plane. The dashed vertical line indicates the location of the photon sphere at $r=r_{\rm ps}$, where the vector field exhibits a topological defect, while the event horizon at $r=r_h$ bounds the physical region. The nontrivial winding of $\mathbf{n}$ around $r_{\rm ps}$ signals a non-vanishing topological charge, confirming the existence of a standard unstable photon sphere outside the event horizon. The total topological charge is $\mathcal{Q}_{\rm tot}=-1$, in agreement with the general result for black hole space-times.}
		\label{fig:vector}
\end{figure}

Fig.~\ref{fig:vector} displays the normalized vector field
$\mathbf{n}$ in the $(r,\theta)$ plane. The photon sphere appears as a topological defect at $r=r_{\rm ps}$, while the event horizon at $r=r_h$ bounds the physical region. The winding of the vector field around the defect yields $\mathcal{Q}_{\rm tot}=-1$, in agreement with the general result for black hole space-times. This negative topological charge is characteristic of an unstable photon sphere, which is precisely what is expected for the outermost light ring in a black hole geometry.
	
\subsection{Topological classification in thermodynamic parameter space}

We now apply the thermodynamic topological framework of Ref.~\cite{Wei:2022dzw}, in which black hole states are identified as zeros of a vector field defined over an extended thermodynamic parameter space. Within this approach, each thermodynamic configuration corresponds to a point in this space, and the zeros of the vector field encode the equilibrium states of the system. The associated winding numbers characterize the local stability of each branch and, when summed over all zeros, define a global topological number for the system.

First, we rewrite the mass function \eqref{eq:M} and the corresponding Hawking temperature \eqref{eq:T} in terms of the horizon radius $r_h$ as
\begin{equation}
	M(r_h)=
	\frac{1}{6}
	\left(
	8\pi P r_h^3
	+\frac{q^4}{(q+r_h)^3}
	+r_h
	\left[
	3+\frac{3q^2}{(q+r_h)^2}
	\right]
	\right),
\end{equation}
and
\begin{equation}
	T(r_h)=
	2 P r_h
	+
	\frac{
		q^4+4q^3 r_h+5q^2 r_h^2+4q r_h^3+r_h^4
	}{
		4\pi r_h (q+r_h)^4
	}.
\end{equation}
respectively. Since the entropy satisfies $S=\pi r_h^2$, there exists a one-to-one correspondence between $r_h$ and $S$. Working directly with $r_h$ therefore represents a smooth reparametrization of the thermodynamic state space, which simplifies the analysis while preserving the underlying physical content. In particular, this change of variables does not affect the associated topological invariants, ensuring that the conclusions drawn from the topological method remain unchanged.

Extrema of the temperature are determined by the condition
\begin{equation}
	\frac{dT}{dr_h}=0,
\end{equation}
which identifies the points where the temperature reaches local maxima or minima as a function of the horizon radius. These extrema play a crucial role in the thermodynamic stability analysis of the system.

A potential van der Waals--type critical point would additionally require the inflection point condition
\begin{equation}
	\frac{d^2T}{dr_h^2}=0,
\end{equation}
which signals the onset of critical behavior associated with a second-order phase transition.

Combining these two conditions leads to the following algebraic equation:
\begin{equation}
	2 q^2 (2q-3r_h) r_h^3 + (q+r_h)^6 = 0.
	\label{criticaleq}
\end{equation}
Eq.~\eqref{criticaleq} admits no solution for $r_h>0$ and $q\ge0$. Therefore, the system does not exhibit any van der Waals--type critical behavior. This result highlights an important distinction from many other charged AdS black hole solutions, which typically display a small/large black hole phase transition analogous to the liquid--gas transition in van der Waals fluids. In the present case, the absence of real, positive solutions indicates that no such critical point exists in the parameter space.

\begin{figure}[t!]
	\centering
	\includegraphics[scale=0.4]{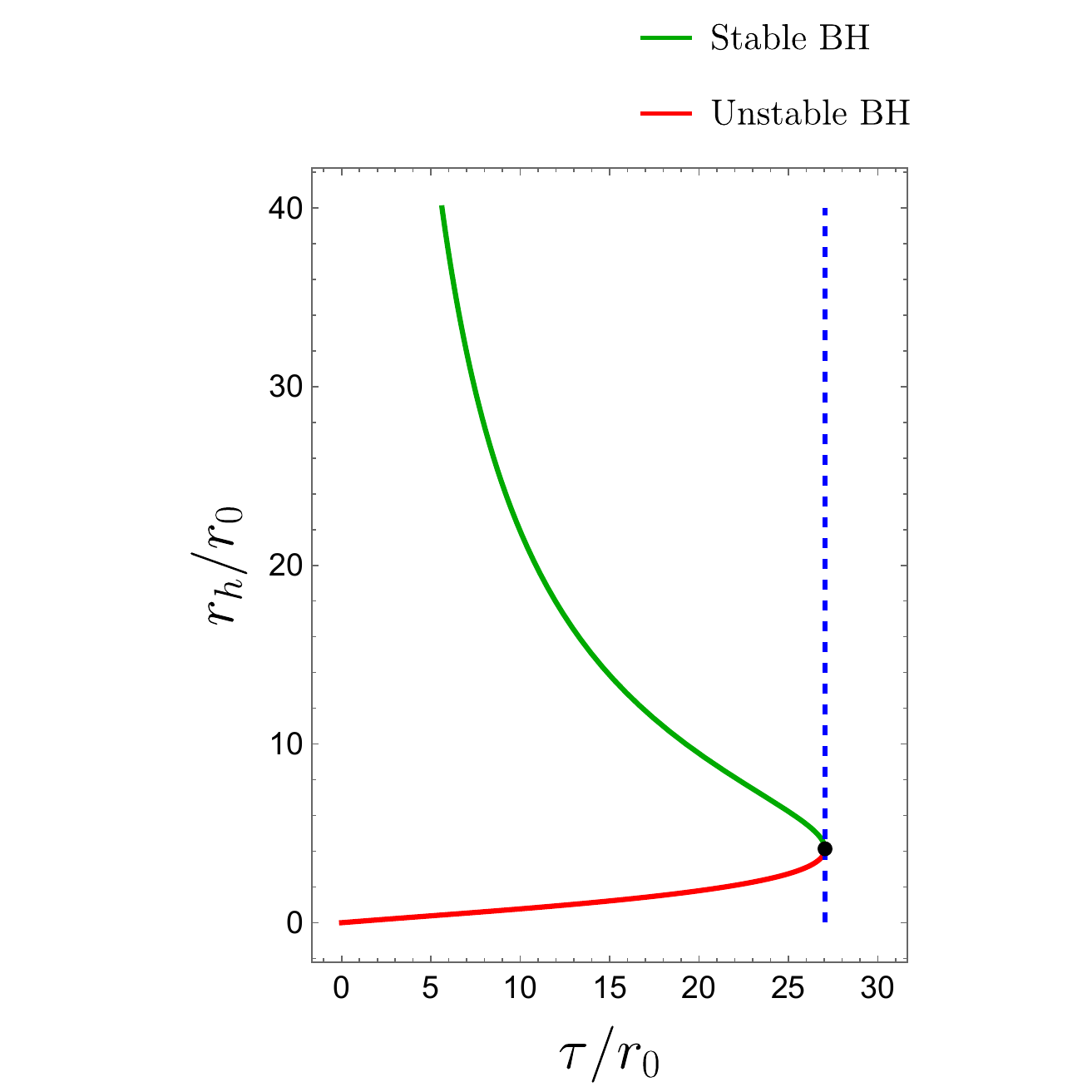}
	\caption{Dimensionless inverse temperature $\tau/r_0$ as a function of the dimensionless horizon radius $r_h/r_0$ for fixed values of the parameters. The curve exhibits a single extremum separating a locally unstable branch ($C_P<0$) from a locally stable one ($C_P>0$). For a fixed value of $\tau$, the intersections with a vertical line correspond to the on-shell black hole states satisfying Eq.~(\ref{onshell}). The absence of multiple extrema confirms that no van der Waals–type phase transition occurs.}
	\label{fig:fasesbh}
\end{figure}

The temperature possesses only a single extremum, which separates a locally unstable branch ($C_P<0$) from a locally stable one ($C_P>0$). This behavior reflects the change in the sign of the heat capacity and therefore signals a transition between thermodynamic instability and stability. This structure is illustrated in Fig.~\ref{fig:fasesbh}, where the dimensionless inverse temperature is plotted as a function of the horizon radius. For a fixed external parameter $\tau$, the intersections of a vertical line with the curve correspond to on-shell black hole states, belonging respectively to the unstable and stable branches.

To implement the thermodynamic topology, we introduce the generalized off-shell free energy
\begin{equation}
	\mathcal{F}(r_h,\tau)
	=
	M(r_h)
	-\frac{\pi r_h^2}{\tau},
\end{equation}
where $\tau$ is an auxiliary inverse temperature parameter. This quantity allows us to extend the thermodynamic description beyond equilibrium configurations, providing a convenient framework to analyze the global structure of the phase space and identify the relevant topological features of the system.

The thermodynamic vector field in the $(r_h,\theta)$ plane is defined as
\begin{equation}
	\boldsymbol{\Psi}
	=
	\left(
	\Psi_{r_h},
	\Psi_{\theta}
	\right),
\end{equation}
where each component encodes information about the thermodynamic behavior of the system in the extended parameter space. Explicitly, the components are given by
\begin{equation}
	\Psi_{r_h} 
	=
	4\pi P r_h^2
	+
	\frac{
		q^4+4q^3 r_h+5q^2 r_h^2+4q r_h^3+r_h^4
	}{
		2(q+r_h)^4
	}
	-
	\frac{2\pi r_h}{\tau},
\end{equation}
\begin{equation}
	\Psi_{\theta}
	=
	-\cot\theta\,\csc\theta.
\end{equation}
respectively. The zeros of the vector field $\boldsymbol{\Psi}$ correspond to equilibrium configurations and are determined by the condition
\begin{equation}
	\Psi_{r_h}=0,
\end{equation}
since $\Psi_{\theta}$ vanishes only at the poles. This condition reproduces the on-shell relation
\begin{equation}
	T(r_h)=\frac{1}{\tau},
	\label{onshell}
\end{equation}
which establishes the connection between the auxiliary parameter $\tau$ and the physical Hawking temperature. Therefore, physical black hole states are identified with the zeros (topological defects) of the vector field, providing a geometric and topological interpretation of thermodynamic equilibrium.

\begin{figure}[t!]
	\centering
	\includegraphics[width=\columnwidth]{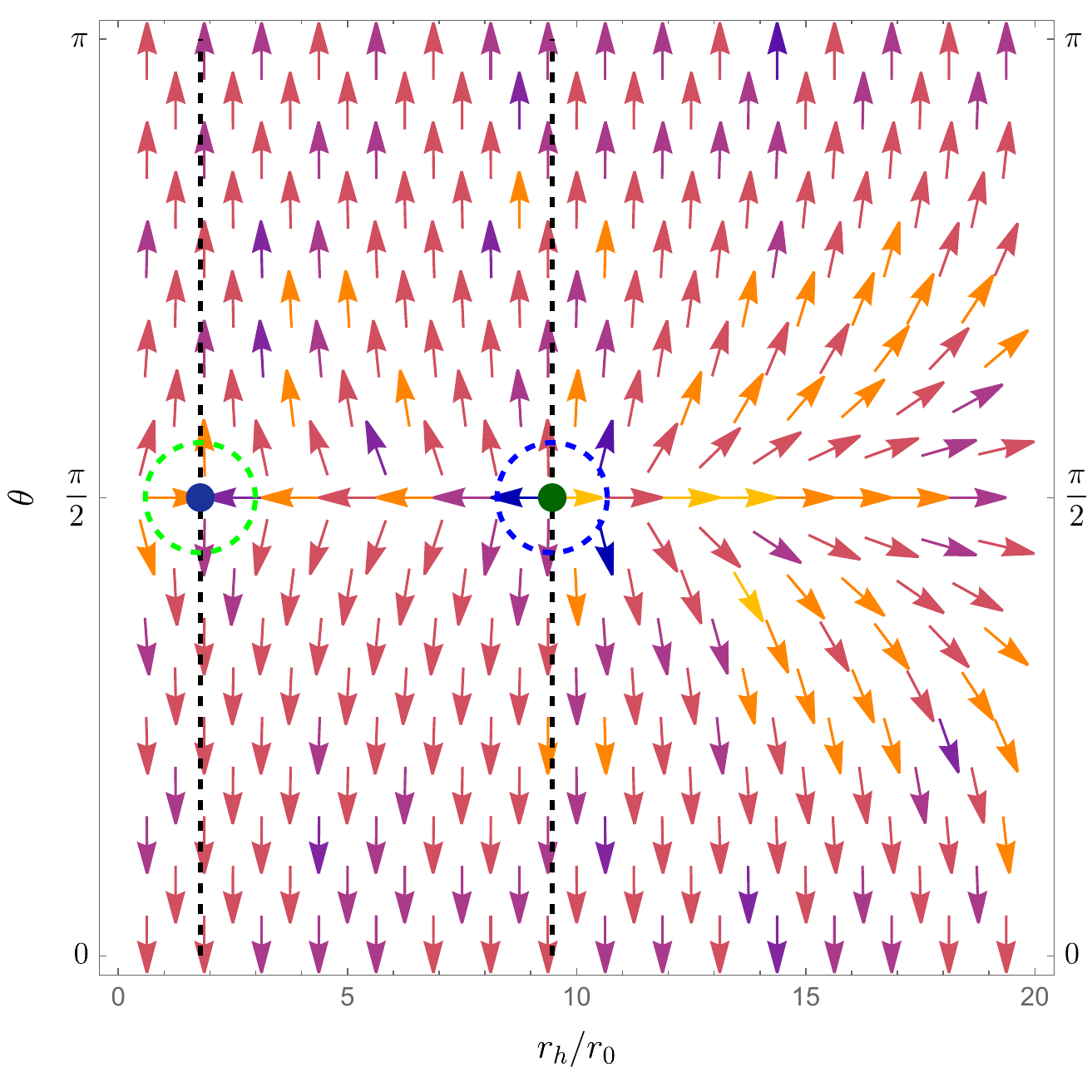}
	\caption{Normalized thermodynamic vector field $\boldsymbol{\Psi}/|\boldsymbol{\Psi}|$ in the $(r_h/r_0,\theta)$ plane for fixed $\tau$. Zeros of the vector field correspond to on-shell black hole states. The winding orientation around each defect determines its topological charge: the stable branch carries winding number $+1$, while the unstable branch carries $-1$.}
		\label{fig:vectortermo}
\end{figure}

For a fixed value of $\tau$, Eq.~\eqref{onshell} admits two positive solutions, associated with the unstable and stable branches identified in Fig.~\ref{fig:fasesbh}. These solutions correspond to distinct thermodynamic phases of the black hole. The behavior of the normalized vector field around these zeros is displayed in Fig.~\ref{fig:vectortermo}, where the local structure of the flow reveals the nature of each configuration. The orientation of the vector field in the neighborhood of each zero determines the corresponding winding number of the defect.

We find that one zero carries winding number $+1$, corresponding to the locally stable black hole branch, while the other carries winding number $-1$, associated with the unstable branch. As a consequence, the total (global) topological number is given by $\mathcal{W}=1-1=0$. This result indicates that the thermodynamic phase structure of the present solution belongs to the same topological class as the Reissner--Nordström black hole.

The absence of additional zeros confirms that no van der Waals--type critical behavior arises in this system, in agreement with the previous analysis. Instead, the thermodynamic configuration space forms a single, globally connected topological sector characterized by $\mathcal{W}=0$. Such a topological classification provides a robust and largely model-independent framework to distinguish different black hole families based on their thermodynamic stability properties and phase structure.

\section{Conclusion}\label{sec:conclu}

In this work we have provided a consistent matter source for a previously proposed static and spherically symmetric black hole geometry. By identifying the deformation parameters with a magnetic charge, we were able to reduce the parameter space and reconstruct the space-time from a well defined variational principle involving nonlinear electrodynamics non-minimally coupled to a scalar field. Within this framework, we found a one-parameter family of electromagnetic Lagrangians, which naturally includes the linear Maxwell case as a particular limit. This construction highlights that non-minimal scalar--electromagnetic couplings constitute a powerful and flexible mechanism for generating physically motivated black hole solutions.

The horizon analysis revealed the existence of a critical magnetic charge separating black hole configurations from horizonless geometries, thus establishing a clear boundary in the parameter space. We also computed the photon sphere and the corresponding shadow radius, allowing us to confront the model with observational constraints from the Event Horizon Telescope for Sagittarius~A*. These bounds restrict the allowed values of the magnetic charge while maintaining consistency with current observational data. In particular, we found that $Q < 1.12$ at $1\sigma$ and $Q < 1.48$ at $2\sigma$, providing a direct and quantitative observational test of the model.

Within extended thermodynamics, the black hole mass satisfies the first law and the corresponding Smarr relation, ensuring the internal consistency of the thermodynamic description. The system exhibits a Hawking--Page phase transition between the thermal AdS background and the black hole configuration, together with a single change of local stability associated with the sign of the heat capacity. However, it does not display van der Waals--type criticality. The absence of additional extrema in the temperature confirms that the thermodynamic structure consists of a single connected sector without the small/large black hole phase transition commonly found in other charged AdS solutions. This behavior is reminiscent of the Reissner--Nordström--AdS black hole, although the presence of the scalar coupling enriches the interpretation of the thermodynamic potentials and their physical origin.

From a topological perspective, the photon sphere appears as a topological defect with total charge $\mathcal{Q}_{\text{tot}}=-1$, as expected for black hole space-times possessing an event horizon. In the thermodynamic parameter space, the zeros of the off-shell vector field correspond to stable and unstable branches characterized by winding numbers $+1$ and $-1$, respectively, yielding a global topological number $\mathcal{W}=0$. This result places the present solution in the same thermodynamic topological class as the Reissner--Nordström black hole. Altogether, the combination of shadow constraints, thermodynamic analysis, and topological classification provides a comprehensive characterization of the solution, highlighting its consistency with both fundamental theoretical expectations and current observational data.

This work pinpoints the versatility and robustness of the non-minimal coupling formalism between nonlinear electrodynamics and scalar fields to describe and reinterpret solutions introduced within GR. In fact, the results presented here illustrate how such couplings can be used to construct consistent matter sources for geometries that were previously introduced at a phenomenological level, thereby providing a deeper theoretical foundation for these space-times. Furthermore, by allowing for richer interactions between matter fields and geometry, it opens the door to a wider class of black hole configurations with potentially observable deviations from standard scenarios. In this sense, this framework offers a promising avenue to connect modified matter sectors with precision measurements in strong gravity environments, thereby contributing to a deeper understanding of both classical and quantum aspects of gravitation.

Several promising directions for future investigation are now open. In particular, it would be natural to extend the present analysis to rotating black hole solutions, where the interplay between angular momentum, scalar fields, and nonlinear electromagnetic effects may lead to qualitatively new features in both the geometry and the associated observables. Another important avenue concerns the study of dynamical space-times, including gravitational collapse and black hole formation, in order to understand how these non-minimal couplings influence the evolution and stability of compact objects. Moreover, confronting predictions of this framework with data from current and future observational facilities—such as next-generation very long baseline interferometry experiments and gravitational wave detectors—would allow for increasingly stringent tests of the model. Such studies would help to place tighter constraints on the parameter space and assess the overall viability of non-minimally coupled NED scenarios, potentially providing a direct window into beyond-Maxwell electromagnetic interactions in strong gravity regimes.

\section*{Acknowledgements}
	
MER thanks Conselho Nacional de Desenvolvimento Cient\'ifico e Tecnol\'ogico - CNPq, Brazil, for partial financial support. This study was financed in part by the Coordena\c{c}\~{a}o de Aperfei\c{c}oamento de Pessoal de N\'{i}vel Superior - Brasil (CAPES) - Finance Code 001.
FSNL acknowledges support from the Funda\c{c}\~{a}o para a Ci\^{e}ncia e a Tecnologia (FCT) Scientific Employment Stimulus contract with reference CEECINST/00032/2018, and funding through the research grant UID/04434/2025.
DRG is supported by the Agencia Estatal de Investigación Grant Nos. PID2022-138607NB-I00 and CNS2024-154444, funded by MICIU/AEI/10.13039/501100011033 (Spain).

	

	
\end{document}